\documentclass[sn-nature]{sn-jnl}

\usepackage{graphicx}%
\usepackage{multirow}%
\usepackage{amsmath,amssymb,amsfonts}%
\usepackage{amsthm}%
\usepackage{mathrsfs}%
\usepackage[title]{appendix}%
\usepackage{xcolor}%
\usepackage{textcomp}%
\usepackage{manyfoot}%
\usepackage{pdfpages}%
\usepackage{booktabs}%
\usepackage{algorithm}%
\usepackage{algorithmicx}%
\usepackage{algpseudocode}%
\usepackage{listings}%
\usepackage{subcaption}%
\usepackage[left]{lineno}%

\theoremstyle{thmstyleone}%
\theoremstyle{thmstyletwo}%

\theoremstyle{thmstylethree}%

\begin{document}


{\title[Jaynes Machine: The universal microstructure of deep neural networks]{Jaynes Machine: The universal microstructure of deep neural networks}}

\author[a*]{Venkat Venkatasubramanian\footnote{Corresponding author: venkat@columbia.edu}}

\affil[a]{Complex Resilient Intelligent Systems Laboratory, Department of Chemical Engineering, Columbia University, New York, NY 10027, U.S.A}

\author[b]{N. Sanjeevrajan}
\affil[b]{Department of Material Engineering, Indian Institute of Technology-Madras, Chennai 600020, India}

\author[c]{Manasi Khandekar}
\affil[c]{Department of Electrical Engineering, Columbia University, New York, NY 10027, U.S.A}


\maketitle

\begin{abstract}

We present a novel theory of the microstructure of deep neural networks. Using a theoretical framework called \textit{statistical teleodynamics}, which is a conceptual synthesis of statistical thermodynamics and potential game theory, we predict that all highly connected layers of deep neural networks have a universal microstructure of connection strengths that is distributed lognormally ($LN (\mu, \sigma$)). Furthermore, under ideal conditions, the theory predicts that $\mu$ and $\sigma$ are the same for all layers in all networks. This is shown to be the result of an arbitrage equilibrium where all connections compete and contribute the same effective utility towards the minimization of the overall loss function. These surprising predictions are shown to be supported by empirical data from six large-scale deep neural networks in real life. We also discuss how these results can be exploited to reduce the amount of data, time, and computational resources needed to train large deep neural networks. 

\end{abstract}

\section{Design of Optimal Teleological Networks}

We consider deep neural networks~\cite{lecun2015deeplearning} to be an example of teleological networked systems that are specifically designed to achieve some goal(s) in uncertain operating environments. Whether human-engineered or naturally evolved, such networked systems exist to deliver desired performance under challenging conditions. For example, supply chains exist to deliver goods and services, and the Internet to facilitate communication robustly.  Similarly, in nature, metabolic networks and ecological food webs exist to support life in myriad ways. This goal-driven feature is an integral aspect of their structure, function, and optimal design. \\

Modeling and analyzing the structure of networked goal-driven systems is an important step toward understanding their design and performance. This is crucial because connection strengths determine the interactions between the agents in the network, and the interactions cumulatively determine the overall performance of the system. In general, network performance depends on two critical design metrics of the system: efficiency and robustness~\cite{venkat2004spontaneous, venkat2007darwin}. We use the term efficiency broadly as a measure of the effectiveness of the network configuration to perform its functions and meet a target level of performance. In airline networks, for example, this would be efficient and safe transportation of passengers. In deep neural networks, this is the efficient minimization of the loss function.  By robustness, we mean the extent to which the network is able to meet the performance target despite variations in its operating environment. Efficiency and robustness are often conflicting objectives to satisfy. Engineers are instinctively aware of the trade-offs they need to make to obtain a satisfactory overall performance that would have to include both efficiency and robustness requirements. Furthermore, these have to be accomplished under cost constraints, which occur in the form of constraints in energy, materials, computational power, money, time, etc. \\

In this broad perspective of goal-driven networked agents, we consider the design of a deep neural network in its most generic and essential form and pose the problem as follows. Given a set of performance specifications and cost constraints, what is the optimal microstructure of a network that can deliver the target performance in a wide variety of uncertain operating environments? \\

We address this general formulation by building on our prior work on network design~\cite{venkat2004spontaneous, venkat2006entropy, venkat2007darwin, venkat2017book, venkat2011supplychain}. We consider a large deep neural network with $L$ layers of neurons. Let layer $l$ have $N^l$ neurons that are connected to the neurons in layer $l-1$ using $M^l$ connections. To benefit from statistical properties, we assume that $M^l$ is of the order of millions. These connections have weights, which can be positive or negative, that determine the strength of the connections. We scaled all the weights to the range 0 to 1, which is divided into $m$ bins. So, any given connection belongs to one of the $m$ bins. The strength of connection of a neuron $i$ in layer $l$ that is connected to a neuron $j$ in layer $l-1$, and belonging to the bin $k$, is denoted by $w_{ijk}^{l}$. The number of connections in bin $k$ for layer $l$ is given by $M_k^l$ with the constraint $M^l = \sum_{k=1}^{m} M_k^l$. The total budget for weights is constrained by $W^l = \sum_{k=1}^{m} M_k^l \mid w_{ijk}^l \mid$.\\

Our deep neural network is teleological. That is, it was human-engineered or naturally evolved to meet certain goals and deliver certain performance targets efficiently and robustly. In this context, efficiency is a measure of how effectively the network minimizes the loss function with minimal use of resources. For example, building and maintaining connections incur costs such as computing power, time, energy, etc. One would like the network to meet its performance target of making accurate predictions with minimal use of such resources. Similarly, by robustness, we mean the ability to deliver the performance target despite variations in its operating environment, such as making accurate predictions in \textit{test} datasets that are different from its \textit{training} datasets.  \\

\section{Statistical Teleodynamics, Population Games, and Arbitrage Equilibrium}

In a typical deep neural network training regimen using gradient descent, the backpropagation algorithm gently nudges all the connections to modify their weights in such a way that the overall loss function is minimized over many iterations and over many datasets. One can equivalently model the same process as the self-organizing competitive dynamics among the connections to modify their weights in such a way that the overall loss function is minimized iteratively over many datasets. We formulate this approach by building on our previous work using a theoretical framework called \textit{ statistical teleodynamics}~\cite{venkat2004spontaneous, venkat2006entropy, venkat2007darwin, venkat2015howmuch, venkat2017book}. It is a synthesis of the central concepts and techniques of population games theory with those of statistical mechanics towards a unified theory of emergent equilibrium phenomena and pattern formation in active matter.\\

In population games, one is interested in the prediction of the final outcome(s) of a large population of goal-driven agents competing dynamically to increase their respective utilities. In particular, one would like to know whether such a game would lead to an equilibrium outcome~\cite{easley2010networks, sandholm2010population}. For some population games, one can identify a single scalar-valued global function, called a {\em potential} $\phi(\boldsymbol{x})$ (where $\boldsymbol{x}$ is the state vector of the system) that captures the necessary information about the utilities of the agents. The gradient of the potential is the utility. Such games are called {\em potential games}~\cite{rosenthal1973class,sandholm2010population, easley2010networks, monderer1996potential}. A potential game reaches strategic equilibrium, called \textit{Nash equilibrium}, when the potential $\phi(\boldsymbol{x})$ is maximized. Furthermore, this equilibrium is unique if $\phi(\boldsymbol{x})$ is strictly concave (ie, $\partial^2 \phi /\partial^2 x < 0$)~\cite{sandholm2010population}. \\

Therefore, an agent's utility, $h_k$, in state $k$ is the gradient of potential $\phi(\boldsymbol{x})$, i.e.,
\begin{equation}
{h}_k(\boldsymbol{x})\equiv {\partial \phi(\boldsymbol{x})}/{\partial x_k}
\label{eq:utility}
\end{equation}
where $x_k=N_k/N$, $\boldsymbol{x}$ is the population vector, {$N_k$ is the number of agents in state $k$, and $N$ is the total number of agents}. By integration (we replace partial derivative with total derivative because ${h}_k(\mathbf{x})$ can be reduced to ${h}_k(x_k)$), we have 
\begin{eqnarray}
\phi(\boldsymbol{x})&=&\sum_{k=1}^m\int {h}_k(\boldsymbol{x}){d}x_k \label{eq:potential}
\end{eqnarray}

where $m$ is the total number of states. \\

To determine the maximum potential, one can use the method of  Lagrange multipliers with $\mathscr{L}$ as the Lagrangian and $\lambda$ as the Lagrange multiplier for the constraint $\sum_{k=1}^mx_k=1$:

\begin{equation}
\mathscr{L}=\phi+\lambda(1-\sum_{k=1}^mx_k)
\label{eq:lagrangian}
\end{equation}\\
If there are other constraints, they can be accommodated similarly~\cite{venkat2015howmuch}.\\

In equilibrium, all agents enjoy the same utility, that is, $h_k = h^*$. It is an \textit{arbitrage equilibrium} \cite{kanbur2020occupational} where the agents no longer have any incentive to switch states, as all states provide the same utility $h^*$. Thus, the maximization of $\phi$ and $h_k = h^*$ are equivalent when the equilibrium is unique (i.e., $\phi(\boldsymbol{x})$ is strictly concave \cite{sandholm2010population}). The former stipulates it from the \textit{top-down, system perspective} whereas the latter is the \textit{ bottom-up, agent} perspective. Thus, this formulation exhibits a \emph{duality} property. \\

We use this formalism to model the self-organizing competitive dynamics among the connections in a deep neural network. We define the \textit{effective utility}, $h_{ijk}^l$, introduced in Eq. \ref{eq:utility}, for a connection with a weight of $w_{ijk}^l$ in layer $l$. The effective utility is a measure of the contribution that this connection makes toward the network-wide reduction in the loss function (i.e., the efficiency component) in a robust manner. The efficiency and robustness metrics, the reader may recall, capture the teleological objective of the network design, which is to deliver certain performance targets in uncertain environments. In the context of deep neural networks, the performance target is to minimize the overall loss function in a robust manner, i.e., for a wide variety of test datasets.  In this perspective, the goal of every neuron is to stay connected with other neurons so that it can process, send, and receive information efficiently under different conditions to minimize the loss function. The more connections of varying weights it has, the more robust its membership in the network against the loss of connections and/or neurons. To accomplish this, the connections compete with each other to provide a more effective utility. i.e., more net benefit towards the goal of minimizing the loss function. Thus, the effective utility of a connection is a benefit-cost trade-off function. It is the net benefit contributed by a connection after accounting for the costs of maintenance and competition, as we discuss below. \\

Thus, the effective utility $h_{ijk}^l$ is made up of three components, 
\begin{equation}
     h_{ijk}^l  = u_{ijk}^l - v_{ijk}^l - z_{ijk}^l
     \label{eq:utility-abc}
\end{equation}

where $u_{ijk}^l$ is the utility derived from the strength of the connection, $v_{ijk}^l$ is the cost or disutility of maintaining such a connection, and $z_{ijk}^l$ is the disutility of competition among connections. Disutilities are costs to be subtracted from the benefit $u_{ijk}^l$. \\

Now, in general, as the strength of the connection $w_{ijk}^{l}$ grows, the marginal utility of its contribution diminishes. This diminishing marginal utility is a commonly found occurrence for many resources and is normally modeled as a logarithmic function. Therefore, the utility $u_{ijk}^l$ derived from this can be written as

\begin{equation}
     u_{ijk}^l  = \alpha^l \ln \mid w_{ijk}^l \mid
     \label{eq:utility-alpha}
\end{equation}
where $\mid w_{ijk}^l \mid$ signifies that  $u_{ijk}^l$ depends only on the absolute magnitude and not on the sign of the weight, and $\alpha^l$ is a parameter. \\

But, as noted, this benefit comes with a cost, as building and maintaining connections are not free. As Venkatasubramanian~\cite{venkat2017book} has shown, most benefit-cost trade-offs in real life are in the form of an inverted-U curve. The simplest model of this behavior is a quadratic function~\cite{venkat2017book}, and so we have $v_{ijk}^l = \beta^l (\ln \mid w_{ijk}^l \mid)^2$, such that
\begin{equation}
     u_{ijk}^l - v_{ijk}^l  = \alpha^l \ln \mid w_{ijk}^l \mid - \beta^l (\ln \mid w_{ijk}^l \mid) ^2
     \label{eq:utility-alpha-beta}
\end{equation} where $\beta^l$ is another parameter. \\

As more and more connections accumulate in the same bin (that is, having the same weight), each new connection is less valuable to the neuron in generating utility. Thus, a neuron would prefer the connections to be distributed over all the bins. This is enforced by the cost term $z_{ijk}^l$. Appealing to diminishing marginal utility again~\cite{venkat2017book}, we model this as $\gamma^l \ln M_k^l$, where $\gamma^l$ is another parameter. \\

 Therefore, the effective utility $h_{ijk}^l$ is given by 
\begin{equation*}
      h_{ijk}^l  = \alpha^l \ln \mid w_{ijk}^l \mid - \beta^l (\ln \mid w_{ijk}^l \mid) ^2 -\gamma^l \ln M_k^l
\end{equation*}\\

We can let $\gamma^l =1$ without any loss of generality and rewrite the equation as 

\begin{equation}
      h_{ijk}^l  = \alpha^l \ln \mid w_{ijk}^l \mid - \beta^l (\ln \mid w_{ijk}^l \mid) ^2 - \ln M_k^l
\label{eq:utility-alpha-beta-gamma}
\end{equation}\\

We wish to point out that the structure of this model is similar to the one we proposed in modeling the Income Game using the statistical teleodynamic framework~\cite{venkat2015howmuch, venkat2017book}. In fact, that is the inspiration for the model proposed here. \\

So, all connections compete with each other to increase their effective utilities ($h_{ijk}^l$) to help reduce the overall loss function in a robust manner. They do this by switching from one state to another by dynamically changing the weights $w_{ijk}^l$, depending on the local gradient of $h_{ijk}^l$, in a manner similar to gradient descent. One of the important results in potential game theory is that this competitive dynamics will result in a Nash equilibrium where the potential $\phi(x)$ is maximized. At equilibrium, all agents enjoy the same utility – that is, $h_{ijk}^l = h^{l*}$ for all $i, j$ and $k$. This is an \textit{arbitrage equilibrium} as all agents have the same utility, thereby removing any incentive to switch states. \\

Using Eq. \ref{eq:utility-alpha-beta-gamma} in Eq. \ref{eq:potential}, we have 
 
\begin{equation}
\phi(\mathbf{x})^l=\phi_u^l + \phi_v^l + \phi_z^l + \text{constant}\label{pay_potential}
\end{equation}
where
\begin{eqnarray}
\phi_u^l&=\alpha^l \sum_{k=1}^mx_k^l\ln \mid w_{ijk}^l \mid\\
\phi_v^l&=-\beta^l \sum_{k=1}^mx_k^l(\ln \mid w_{ijk}^l \mid)^2\\
\phi_z^l&=\frac{1}{M^l} \ln \frac{M^l!}{\prod_{k=1}^m(M^lx_k^l)!}
\label{fair_potential}
\end{eqnarray}

where $x_k^l=M_k^l/M^l$ and we have used Stirling's approximation in equation~\eqref{fair_potential}.\\

We see that $\phi(\mathbf{x})^l$ is strictly concave:
\begin{eqnarray}
{\partial^2 \phi(\mathbf{x})^l}/{\partial x_k^{l2}}=-{1}/{x_k^l}<0
\end{eqnarray}

Therefore, a {\em unique Nash Equilibrium} for this game exists, where $\phi(\mathbf{x})$ is maximized. Using the Lagrangian multiplier approach (Eq. \ref{eq:lagrangian}), we maximize $\phi(\mathbf{x})$ in equations~\eqref{pay_potential}-\eqref{fair_potential} to determine that the equilibrium distribution of the connection weights follows a lognormal distribution, given by \\

\begin{equation}
x_k^l=\frac{1}{\mid w_{ijk}^l \mid \sigma^l \sqrt{2\pi}}\exp\left[-\frac{(\ln \mid w_{ijk}^l \mid - \mu^l)^2}{2\sigma^{l2}} \right]
\label{logn_potential}
\end{equation}\\

where, $\mu^l = \frac {\alpha^l + 1}{2\beta^l}$ and $\sigma^{l} = \sqrt{\frac{1}{2\beta^l}}$. \\

Thus, the theory predicts a surprising and useful result that the microstructure of deep neural networks, i.e., the distribution of connection weights, is lognormal for all highly connected layers. This universality is independent of the size of the network, its architecture, or its application domain. The intuitive explanation is that, in a given layer, all individual connections contribute an effective utility (i.e., a net benefit) toward the overall objective of the network, which is to learn robustly the structure of a complex high-dimensional manifold by minimizing the loss function.  In a large deep neural network, with hundreds of layers and millions of connections in each layer, no connection is particularly unique. No connection is more important than another. Every connection has thousands of counterparts elsewhere, so no one is special. Therefore, there is this inherent symmetry and equality in the microstructure. Hence, they all end up contributing the same effective utility towards that layer's goal of minimizing the loss function as the training progresses. That is why, when training is completed, one reaches the arbitrage equilibrium where all effective utilities are equal in that layer, i.e., $h_{ijk}^l = h^{l*}$ for all $i, j$, and $k$.\\

Furthermore, in the “thermodynamic limit” of extremely large networks, i.e. $L \rightarrow \infty$, $M^l \rightarrow \infty$, and $W^l \rightarrow \infty$, \textit{all} connections in \textit{all} the layers end up making the \textit{same} effective utility contribution, i.e. $h_{ijk}^l = h^{*}$ for all $i,j,k$, and $l$. For this ideal deep neural network, all layers will have a lognormal weight distribution with the \textit{same} $\mu$ and $\sigma$. In other words, $\alpha^l$ and $\beta^l$ are the \textit{same} for all layers. This is the ultimate universal microstructure for ideal deep neural networks. \\

Now, readers familiar with statistical mechanics will recognize the potential component  $\phi_z^l$  as {\em entropy} (except for the missing Boltzmann constant $k_B$). Thus, by maximizing $\phi^l$ in the Lagrangian multiplier formulation, one is equivalently maximizing entropy subject to the constraints specified in the terms $\phi_u^l$ and $\phi_v^l$. Thus, the lognormal distribution is the maximum entropy distribution under these constraints. This connection with entropy reveals an important insight into the robustness property of the network design, as discussed in the next section.\\

The ideal deep neural network is the conceptual equivalent of the ideal gas in statistical thermodynamics. Just as the maximum entropy distribution of energy in statistical thermodynamics is the well-known exponential distribution, called the Boltzmann distribution, we observe that its equivalent in statistical teleodynamics for deep neural networks is the lognormal distribution.\\

\subsection{Optimally Robust Design}
The maximum-entropy design distributes the weights in the network in such a way that it maximizes the uncertainty about a wide variety of future datasets whose nature is unknown, unknowable, and, therefore, uncertain. Thus, in maximum-entropy design, the network is optimized for all potential future environments, not for any particular one. Note that for any particular dataset, one can design a weight distribution such that it will outperform the maximum entropy design with respect to the loss function. However, such a biased network may not perform as well for other datasets, while the maximum entropy distribution-based network is likely to perform better. For instance, if a network is overfitted on a specific dataset, then it might "memorize" these data and hence might not perform that well for other datasets. To prevent this, one uses techniques such as data segmentation, weight regularization, dropout, early stopping, etc. The combined effect of such procedures is to achieve robustness in performance on a wide range of datasets. The goal of such techniques is to accommodate as much variability and as much uncertainty as possible in the test environments. This is exactly what we achieve by maximizing entropy in our theory. Maximizing entropy is the same as maximizing the uncertainty and variability of future datasets. In our theory, this robustness requirement is naturally built in from the very beginning as an integral part of the effective utility and potential function formulation, not as \textit{ad hoc} afterthoughts to prevent overfitting. This is what we mean by \textit{optimally robust design}~\cite{venkat2007darwin}. \\

Thus, an optimally robust deep neural network is a robust prediction engine. It is a maximum entropy machine that learns an efficient and robust model of the target manifold. In the "thermodynamic limit," all networks, such as the Boltzmann Machine, Hopfield network, and so on, are various special instances of this general class.  We call this machine the \textit{Jaynes Machine} in honor of Professor E. T. Jaynes, who elucidated the modern interpretation of the maximum entropy principle in the 1950s ~\cite{jaynes1957information, jaynes1957information2, jaynes1979standonmaxentropy, jaynes1985wheredowego}. \\

\section{Empirical Results}

The predictions of the theory were tested by analyzing the weight distributions in six different deep neural networks. They are (i) BlazePose, (ii) Xception, (iii) BERT-Small, (iv) BERT-Large, (v) Llama-2 (7B), and (vi) LLAMA-2 (13B) \cite{BlazePose2023active, Xception2023active, BERT2023active, LLAMA2023active}. Their salient features are summarized in Table~\ref{tab:table1}. The first two utilize convolution layers, and the other four are based on the transformer architecture~\cite{LLAMA2web,BERTweb,BlazePoseweb}. They are of widely different sizes with respect to the number of parameters and are designed for different application domains. \\

The layer-by-layer weight data for these networks were extracted, normalized between 0 and 1, converted to their absolute magnitudes by dropping the signs, and classified into different bins. For all these networks, some layers had only a few thousand data points (out of the millions or tens of millions in the network), so we did not fit a distribution as statistical measures such as $R^2$ were not good. \\

\begin{table}
    \centering
    \begin{tabular}{c c c c}
    \hline
    Model & Architecture & Parameters size & Application \\ \hline
        BlazePose & Convolution & $2.8\times10^6$ & Computer Vision \\
        Xception & Convolution & $20\times10^6$ & Computer Vision \\
        BERT Small & Transformer & $109\times10^6$ & Natural Language Processing \\
        BERT Large & Transformer & $325\times10^6$ & Natural Language Processing \\
        LLAMA-2 (7B) & Transformer & $7\times10^9$ & Natural Language Processing\\
        LLAMA-2 (13B) & Transformer & $13\times10^9$ & Natural Language Processing \\
    \hline
    \end{tabular}
    \caption{Six deep neural network case studies}
    \label{tab:table1}
\end{table}

The plots show the \textit{size-weighted} distributions (noted as category weight in the y-axis) rather than the weight distribution, since the features are clearer in the former. The category weight is simply the product of the size (i.e., weight) of a category (i.e., bin) and the number of connections in that category. There is a well-known result in statistics \cite{rao1984size} that if the weight distribution is lognormal with $\mu$ and $\sigma$ (i.e., $LN(\mu, \sigma)$), then the size-weighted distribution is also lognormal, $LN(\mu^{\prime}, \sigma^{\prime})$), where $\mu^{\prime} = \mu + \sigma^2$ and $\sigma^{\prime} = \sigma$. Furthermore, since the utility $u_{ijk}^l$ in Eq.~\ref{eq:utility-alpha} is positive (since it is a benefit), and $\ln \mid w_{ijk}^l \mid$ is negative in the range of $0 < \mid w_{ijk}^l \mid < 1$, we have $\alpha^l < 0$, $\mu^l < 0$, and $\mu^{\prime l}< 0$. Similarly, from Eq.~\ref{eq:utility-alpha-beta}, the disutility $v_{ijk}^l$ requires $\beta^l > 0$. 

\begin{figure}[!ht]
    \centering
    \includegraphics[width = 0.9\linewidth]{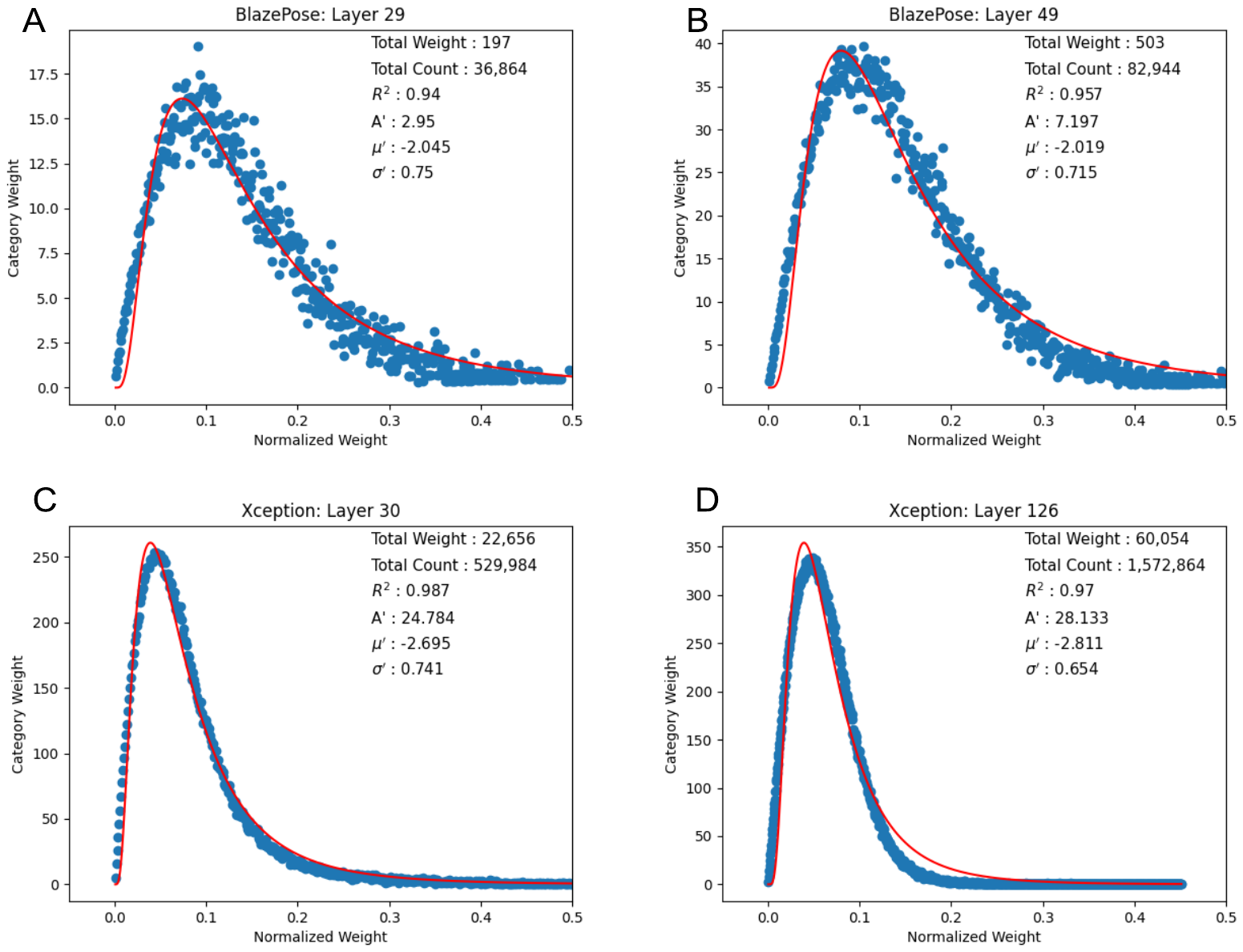}
    \caption{Typical lognormal fitted curves: A \& B - BlazePose; C \& D  - Xception. Blue dots are data, and the red curve is the lognormal fit. The parameters of the fits are also shown.}
    \label{fig:BlazePose-Xception}
\end{figure}

\begin{figure}[H]
    \centering
    \includegraphics[width = 01\linewidth]{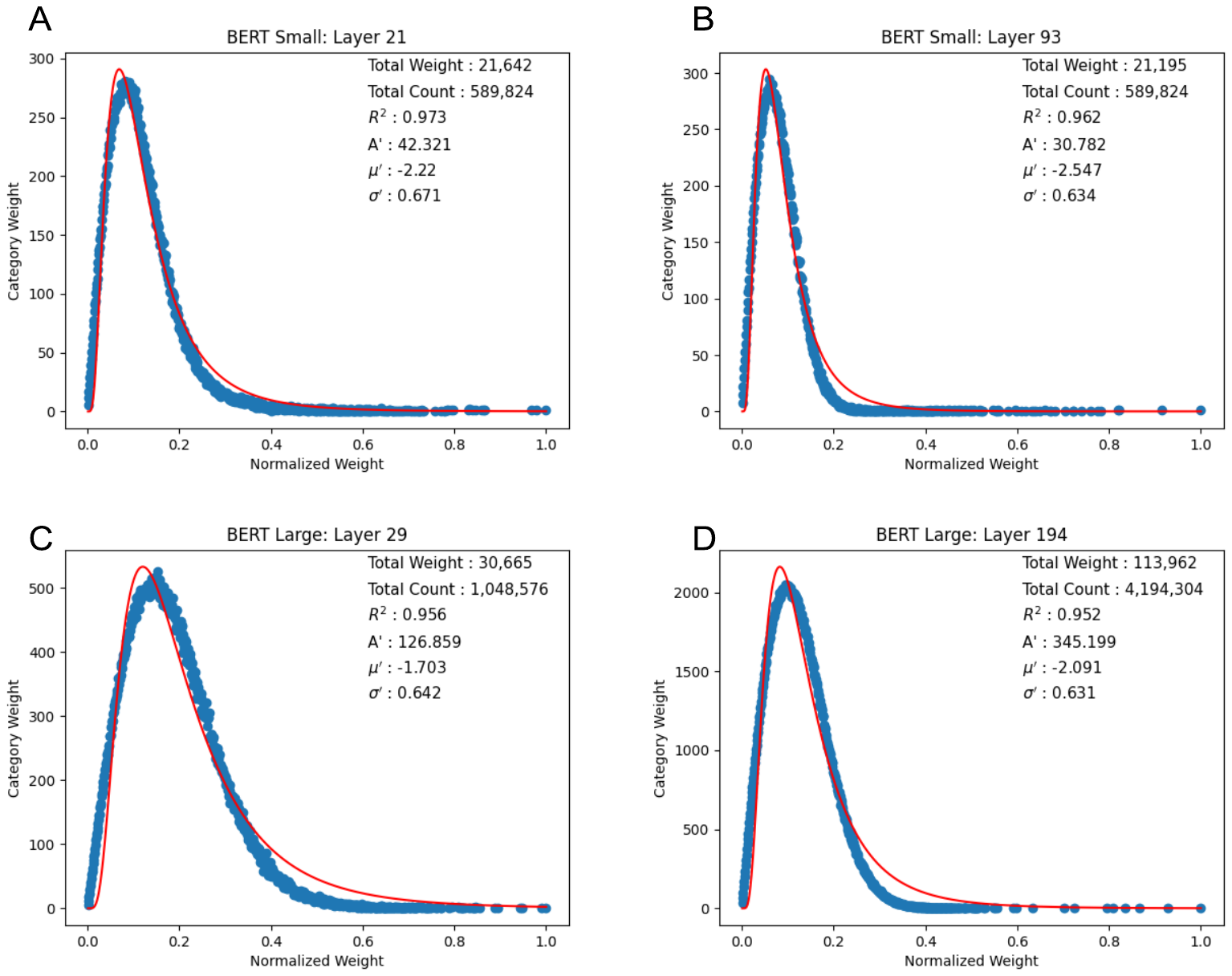}
    \caption{A \& B - BERT-Small; C \& D - BERT-Large}
    \label{fig:BERT}
\end{figure}
\begin{figure}[H]
    \centering
    \includegraphics[width = 01\linewidth]{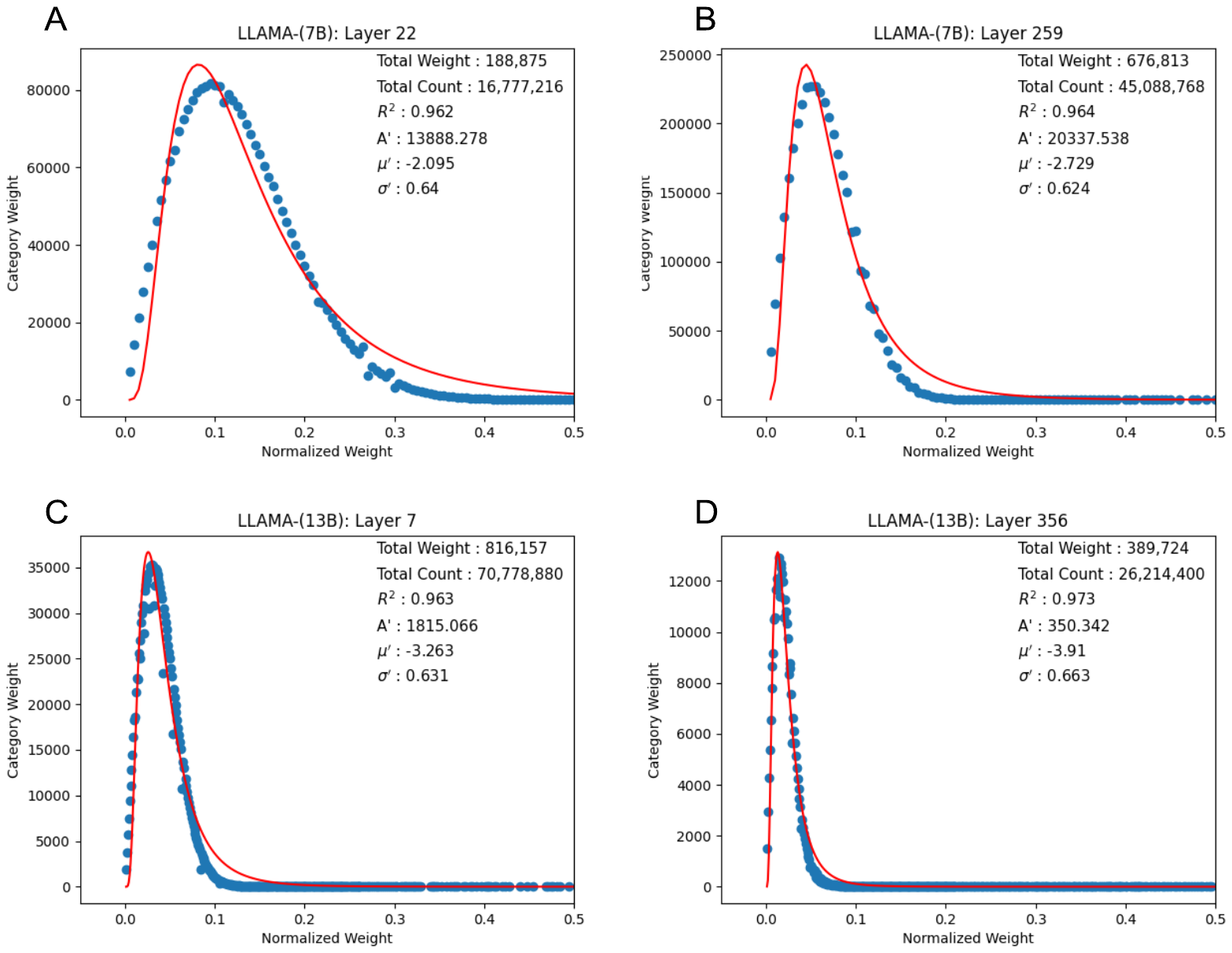}
    \caption{A \& B - Llama-2 7B; C \& D - Llama-2 13B}
    \label{fig:LLAMA-2}
\end{figure}

The lognormal distribution was fitted and tested for the following networks: BlazePose (39 layers), Xception (32 layers), BERT-Small (75 layers), BERT-Large (144 layers), Llama-2 7B (226 layers) and Llama-2 13B (282 layers). Instead of showing the plots for all the 798 layers, which all look pretty similar, we show a much smaller selection of sample distributions in Figs. \ref{fig:BlazePose-Xception}-\ref{fig:LLAMA-2}. We show two typical distributions, one at the beginning and one near the end of the network, for each of the six networks we analyzed. We see that these size-weighted data fit the lognormal distribution very well with high $R^2$ values.  This is typical of all the layers with high connectivity. Although the six networks use different architectures, are of different sizes, and are trained for different applications, we find this surprising universal microstructure. This is an important design feature of these networks that emerges automatically during training. As discussed in Section 2, our theory predicts this universal lognormal microstructure. \\

\begin{table}
    \centering
    \begin{tabular}{c c c c c c}
    \hline
    Model & Layers & $R^2$ & $A^{\prime}$ & $\mu^{\prime}$ & $\sigma^{\prime}$ \\ \hline
        BlazePose & 39 & $0.93 \pm 0.02$ & $3.75 \pm 2.09 $  & $-1.74 \pm 0.52 $ & $1.49 \pm 0.60 $\\
        Xception & 32 & $0.98 \pm 0.01$  & $6.53 \pm 3.64 $  & $-2.87 \pm 0.18 $ & $0.70 \pm 0.05 $\\
        BERT Small & 75 & $0.96 \pm 0.01$  & $66.15 \pm 46.46$  & $-2.47 \pm 0.95 $ & $0.65 \pm 0.02 $\\
        BERT Large & 144 & $0.96 \pm 0.01$  & $44.84 \pm 143.6$  & $-2.37 \pm 0.98 $ & $0.64 \pm 0.01 $\\
        LLAMA-2 (7B) & 226 & $0.97 \pm 0.01$  & $11464 \pm 8170$  & $-2.96 \pm 0.54 $ & $0.66 \pm 0.05 $\\
        LLAMA-2 (13B) & 282 & $0.94 \pm 0.03$  & $1513 \pm 1116$  & $-3.02 \pm 0.53 $ & $0.67 \pm 0.06 $\\
    \hline
    \end{tabular}
    \caption{Average and standard deviation of Lognormal parameters}
    \label{tab:table2}
\end{table}

In the Supplemental Information section, we list the lognormal parameters ($A^{\prime}$, $\mu^{\prime}$, and $\sigma^{\prime}$) for all the highly connected layers (798 of them) for all six case studies. Table~\ref{tab:table2} summarizes the average and standard deviation values of the lognormal distribution parameters for the six case studies. Note that for large networks with $>100$ million connections,  $\sigma^{\prime}$ appears to be nearly constant (around 0.65) for all networks, as seen by its low standard deviation values in Table~\ref{tab:table2}. This implies that $\beta^{\prime}$ is also approximately constant for all networks. Even $\mu^{\prime}$ (and hence $\alpha^{\prime}$) appears to be in a tight range (-2.5 to -3.0) for the different networks. The theory predicts that $\mu^{\prime}$ and $\sigma^{\prime}$ are constants for all networks only in the "thermodynamic limit" of the ideal network. However, we see such a trend even for these nonideal cases.\\

\begin{figure}[H]
    \begin{subfigure}{0.45\linewidth}
        \centering
        \includegraphics[width=\linewidth]{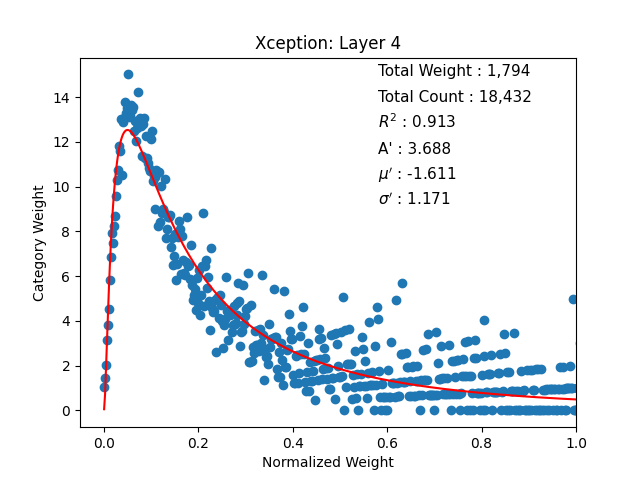}
        \caption{Noisy data in a layer with low number of connections: Xception Layer \#4}
        \label{fig:noisyfig1}
    \end{subfigure}%
    \hfill
    \begin{subfigure}{0.45\linewidth}
        \centering
        \includegraphics[width=\linewidth]{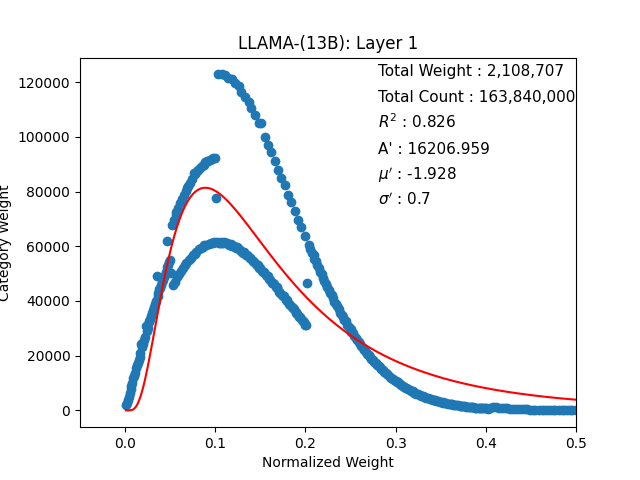}
        \caption{Suboptimal training in a highly connected layer: LLAMA-2 13B - Layer \#1}
        \label{fig:noisyfig2}
    \end{subfigure}
    \caption{Typical problems with low and high number of connections}
    \label{fig:horizontal}
\end{figure}

The number of connections in the 798 layers we studied ranged from 36,864 to  163,840,000. Generally speaking, we find that the more connections a layer has, the better the lognormal fit with higher $R^2$ due to better statistical averaging. We can see in Fig. \ref{fig:noisyfig1} that Layer \#4 of the Xception network, which has only 18,432 connections, has too much noise in the data to fit any distribution well. Therefore, we did not model such layers.\\

However, layers with scores of millions of connections have their own challenges, as they are harder to train, and hence run the risk of suboptimal weight assignments. Recall that, according to the theory, the lognormal distribution emerges only when arbitrage equilibrium is reached. It is possible that these extremely highly connected layers had not quite reached equilibrium when training was stopped. Therefore, naturally, there would be a mismatch between theoretical predictions and empirical observations. We observe this in the Llama-2 (13B) data. Fig. \ref{fig:noisyfig2} shows the size-weighted distribution for Layer \#1, which has over 163 million weights. As we can clearly see, there are elements of the lognormal distribution present, but the fit is not as good as it is for Layer \#7 (see Fig.\ref{fig:LLAMA-2}C), for example, with about 70 million weights. This suggests that Layer \#1 training was suboptimal. It appears from our empirical analysis that layers that have connections in the range of about 1 to 50 million have the right trade-off between better statistical properties and reaching optimal weight distribution.   \\

\section{Conclusions}

Understanding the microstructure of large deep neural networks is of great importance, given their enormous role in many applications. This understanding could lead to practical benefits, such as better design and training algorithms. But equally importantly, it could also lead to better theories about their structure, function, and behavior. Toward that goal, we present a novel theoretical and empirical analysis of the distribution of weights of six different large neural networks. \\

We show both theoretically and empirically that in large neural networks, the final connection strengths (i.e., weights) are distributed lognormally in highly connected layers. This pattern is independent of the architecture or the application domain. We should be able to take advantage of this knowledge to reduce the amount of time, data, and computational resources required to train large deep neural networks effectively. We believe that there are at least three ways in which this knowledge can be utilized to train deep neural network models.\\

First, since we know the final weight distribution is lognormal, we can ‘hot start’ and initialize the weights lognormally instead of randomly. Furthermore, since we have the average values of $A^{\prime}$, $\mu^{\prime}$, and $\sigma^{\prime}$ from these six case studies, they provide us with guidance on the initial weights. This should bring us closer to the finish line when we start. This raises an interesting question: Is this what pre-training is doing for GPTs? Does unsupervised learning transform the initial random distribution of weights into a nearly lognormal distribution, which makes subsequent supervised learning easier? We need further studies to answer such questions.\\

Second, during training using iterative gradient descent, instead of individually tuning millions of weights, we can tune the much smaller number of lognormal parameters. Consider, for example, layer \#7 of Llama-2 (13B), which has about 70 million weights. However, this distribution can be modeled by only three parameters ($A^{\prime}$, $\mu^{\prime}$, and $\sigma^{\prime}$) of the corresponding lognormal distribution. Therefore, we can modify the backpropagation algorithm to tune just these three parameters rather than adjusting 70 million weights. One could do this at least for the initial stages of training and reserve the more resource-consuming fine-tuning of all the weights towards the last stages of training. This kind of hybrid training could result in considerable savings in data, time, and computational resources for large networks. Furthermore, since very large networks struggle to reach optimal allocation of weights (see Fig. 4B), constraining the weight distribution to lognormal in the training iterations, which is the theoretical optimum, minimizes the chances of settling down in suboptimal microstructures. \\

Third, we can design special-purpose hardware where the layers are connected in a lognormal manner with tunable connection strengths. We are currently pursuing all of these opportunities. \\

We wish to emphasize that the spirit of our modeling is similar to that of the ideal gas or the Ising model in statistical mechanics. Just as real molecules are not point-like objects or devoid of intermolecular interactions, as assumed in the ideal gas model, we make similar simplifying assumptions in our model.  The ideal version serves as a useful starting point for developing more comprehensive models. Furthermore, just as real gases do not behave like the ideal gas, we do not expect real-life deep neural networks to behave like their ideal version. That is why it comes as a surprise that the six networks we analyzed come this close to the predictions made for the ideal version.\\

We also stress an important insight revealed by our theory. It is generally viewed that active matter systems such as neural networks are out-of-equilibrium or far-from-equilibrium systems. However, both our theoretical and empirical analyses demonstrate that they are actually in equilibrium, an \textit{arbitrage} equilibrium. Just as systems are in mechanical equilibrium when forces or pressures are equal, in thermal equilibrium when temperatures are equal, or in phase equilibrium when chemical potentials are equal, we have active matter systems in arbitrage equilibrium when effective utilities are equal \cite{venkat2022unified}.\\

The crucial feature of the maximum entropy design, expressed by the lognormal distribution at arbitrage equilibrium, is that the effective utilities of all the connections are equal. This equality reflects a deep sense of balance, harmony, and fairness in network design. This is an elegant solution to the credit assignment problem among millions of connections. One could view this equality as the mathematical criterion of beauty. One cannot help but wonder whether nature has discovered this beautiful secret and exploits it in her evolutionary design of biological brains. 

\vskip6pt

\subsection*{Acknowledgements}
We would like to thank Professor Babji Srinivasan of IIT-Madras for his valuable suggestions.

\subsection*{Author Contributions}

VV: Conceptualization, Theory, Methodology, Formal Analysis, Investigation, Supervision, and Writing; NS: Software development and analysis for the BlazePose, BERT-Small, BERT-Large, Llama-2 (7B), and Llama-2 (13B) networks; MK: Software development and analysis for the Xception network. 

The authors have no conflicts of interest to declare.


\bibliography{sn-bibliography.bib}

\begin{thebibliography}{10}
\expandafter\ifx\csname url\endcsname\relax
  \def\url#1{\burl{#1}}\fi
\expandafter\ifx\csname urlprefix\endcsname\relax\def\urlprefix{URL }\fi
\providecommand{\bibinfo}[2]{#2}
\providecommand{\eprint}[2][]{\url{#2}}
\providecommand{\doi}[1]{\url{https://doi.org/#1}}
\bibcommenthead

\bibitem{lecun2015deeplearning}
\bibinfo{author}{LeCun, Y.}, \bibinfo{author}{Bengio, Y.} \& \bibinfo{author}{Hinton, G.}
\newblock \bibinfo{title}{Deep learning}.
\newblock \emph{\bibinfo{journal}{Nature}} \textbf{\bibinfo{volume}{521}}, \bibinfo{pages}{436--444} (\bibinfo{year}{2015}).

\bibitem{venkat2004spontaneous}
\bibinfo{author}{Venkatasubramanian, V.}, \bibinfo{author}{Katare, S.}, \bibinfo{author}{Patkar, P.~R.} \& \bibinfo{author}{Mu, F.-p.}
\newblock \bibinfo{title}{Spontaneous emergence of complex optimal networks through evolutionary adaptation}.
\newblock \emph{\bibinfo{journal}{Computers \& chemical engineering}} \textbf{\bibinfo{volume}{28}}, \bibinfo{pages}{1789--1798} (\bibinfo{year}{2004}).

\bibitem{venkat2007darwin}
\bibinfo{author}{Venkatasubramanian, V.}
\newblock \bibinfo{title}{A theory of design of complex teleological systems: Unifying the darwinian and boltzmannian perspectives} (\bibinfo{year}{2007}).

\bibitem{venkat2006entropy}
\bibinfo{author}{Venkatasubramanian, V.}, \bibinfo{author}{Politis, D.~N.} \& \bibinfo{author}{Patkar, P.~R.}
\newblock \bibinfo{title}{Entropy maximization as a holistic design principle for complex optimal networks}.
\newblock \emph{\bibinfo{journal}{AIChE journal}} \textbf{\bibinfo{volume}{52}}, \bibinfo{pages}{1004--1009} (\bibinfo{year}{2006}).

\bibitem{venkat2017book}
\bibinfo{author}{Venkatasubramanian, V.}
\newblock \emph{\bibinfo{title}{How Much Inequality Is Fair?: Mathematical Principles of a Moral, Optimal, and Stable Capitalist Society}}  (\bibinfo{publisher}{Columbia University Press}, \bibinfo{year}{2017}).

\bibitem{venkat2011supplychain}
\bibinfo{author}{Shukla, A.}, \bibinfo{author}{Agarwal~Lalit, V.} \& \bibinfo{author}{Venkatasubramanian, V.}
\newblock \bibinfo{title}{Optimizing efficiency-robustness trade-offs in supply chain design under uncertainty due to disruptions}.
\newblock \emph{\bibinfo{journal}{International Journal of Physical Distribution \& Logistics Management}} \textbf{\bibinfo{volume}{41}}, \bibinfo{pages}{623--647} (\bibinfo{year}{2011}).

\bibitem{venkat2015howmuch}
\bibinfo{author}{Venkatasubramanian, V.}, \bibinfo{author}{Luo, Y.} \& \bibinfo{author}{Sethuraman, J.}
\newblock \bibinfo{title}{How much inequality in income is fair?: A microeconomic game theoretic perspective}.
\newblock \emph{\bibinfo{journal}{Physica A: Statistical Mechanics and its Applications}} \textbf{\bibinfo{volume}{435}}, \bibinfo{pages}{120--138} (\bibinfo{year}{2015}).

\bibitem{easley2010networks}
\bibinfo{author}{Easley, D.}, \bibinfo{author}{Kleinberg, J.} \emph{et~al.}
\newblock \emph{\bibinfo{title}{Networks, crowds, and markets}} Vol.~\bibinfo{volume}{8} (\bibinfo{publisher}{Cambridge university press Cambridge}, \bibinfo{year}{2010}).

\bibitem{sandholm2010population}
\bibinfo{author}{Sandholm, W.~H.}
\newblock \emph{\bibinfo{title}{Population games and evolutionary dynamics}}  (\bibinfo{publisher}{MIT press}, \bibinfo{year}{2010}).

\bibitem{rosenthal1973class}
\bibinfo{author}{Rosenthal, R.~W.}
\newblock \bibinfo{title}{A class of games possessing pure-strategy nash equilibria}.
\newblock \emph{\bibinfo{journal}{International Journal of Game Theory}} \textbf{\bibinfo{volume}{2}}, \bibinfo{pages}{65--67} (\bibinfo{year}{1973}).

\bibitem{monderer1996potential}
\bibinfo{author}{Monderer, D.} \& \bibinfo{author}{Shapley, L.~S.}
\newblock \bibinfo{title}{Potential games}.
\newblock \emph{\bibinfo{journal}{Games and economic behavior}} \textbf{\bibinfo{volume}{14}}, \bibinfo{pages}{124--143} (\bibinfo{year}{1996}).

\bibitem{kanbur2020occupational}
\bibinfo{author}{Kanbur, R.} \& \bibinfo{author}{Venkatasubramanian, V.}
\newblock \bibinfo{title}{Occupational arbitrage equilibrium as an entropy maximizing solution}.
\newblock \emph{\bibinfo{journal}{The European Physical Journal Special Topics}} \textbf{\bibinfo{volume}{229}}, \bibinfo{pages}{1661--1673} (\bibinfo{year}{2020}).

\bibitem{jaynes1957information}
\bibinfo{author}{Jaynes, E.~T.}
\newblock \bibinfo{title}{Information theory and statistical mechanics}.
\newblock \emph{\bibinfo{journal}{Physical review}} \textbf{\bibinfo{volume}{106}}, \bibinfo{pages}{620} (\bibinfo{year}{1957}).

\bibitem{jaynes1957information2}
\bibinfo{author}{Jaynes, E.~T.}
\newblock \bibinfo{title}{Information theory and statistical mechanics. ii}.
\newblock \emph{\bibinfo{journal}{Physical review}} \textbf{\bibinfo{volume}{108}}, \bibinfo{pages}{171} (\bibinfo{year}{1957}).

\bibitem{jaynes1979standonmaxentropy}
\bibinfo{author}{Jaynes, E.~T.}
\newblock \bibinfo{title}{Where do we stand on maximum entropy}.
\newblock \emph{\bibinfo{journal}{The maximum entropy formalism}} \bibinfo{pages}{15--118} (\bibinfo{year}{1979}).

\bibitem{jaynes1985wheredowego}
\bibinfo{author}{Jaynes, E.~T.}
\newblock \bibinfo{title}{Where do we go from here?}
\newblock \emph{\bibinfo{journal}{Maximum-Entropy and Bayesian Methods in Inverse Problems}}  (\bibinfo{year}{1985}).

\bibitem{BlazePose2023active}
\bibinfo{author}{Valentin~Bazarevsky, I.~G.}
\newblock \bibinfo{title}{Blazepose: On-device real-time body pose tracking}.
\newblock \emph{\bibinfo{journal}{arXiv:2006.10204v1}}  (\bibinfo{year}{2020}).

\bibitem{Xception2023active}
\bibinfo{author}{Chollet, F.}
\newblock \bibinfo{title}{Xception: Deep learning with depthwise separable convolutions}.
\newblock \emph{\bibinfo{journal}{arXiv}}  (\bibinfo{year}{2017}).

\bibitem{BERT2023active}
\bibinfo{author}{Devlin, J.}, \bibinfo{author}{Chang, M.-W.}, \bibinfo{author}{Lee, K.} \& \bibinfo{author}{Toutanova, K.}
\newblock \bibinfo{title}{Bert: Pre-training of deep bidirectional transformers for language understanding}.
\newblock \emph{\bibinfo{journal}{arXiv}}  (\bibinfo{year}{2019}).

\bibitem{LLAMA2023active}
\bibinfo{author}{Hugo~Touvron, K.~S., Louis~Martin}.
\newblock \bibinfo{title}{Llama 2: Open foundation and fine-tuned chat models}.
\newblock \emph{\bibinfo{journal}{arXiv}}  (\bibinfo{year}{2023}).

\bibitem{LLAMA2web}
\bibinfo{title}{The model parameters for llama, hugging face llama-2}.
\newblock \bibinfo{howpublished}{https://huggingface.co/meta-llama/Llama-2-7b/tree/main}.

\bibitem{BERTweb}
\bibinfo{title}{The model parameters for bert, hugging face llama-2}.
\newblock \bibinfo{howpublished}{https://huggingface.co/docs/transformers/model\_doc/bert}.

\bibitem{BlazePoseweb}
\bibinfo{title}{Medium, blazepose : A 3d pose estimation model}.
\newblock \bibinfo{howpublished}{https://medium.com/axinc-ai/blazepose-a-3d-pose-estimation-model-d8689d06b7c4}.

\bibitem{rao1984size}
\bibinfo{author}{Patil, G.~P.} \& \bibinfo{author}{Rao, C.~R.}
\newblock \bibinfo{title}{Weighted distributions and size-biased sampling with applications to wildlife populations and human families}.
\newblock \emph{\bibinfo{journal}{Biometrics}} \textbf{\bibinfo{volume}{34(2)}}, \bibinfo{pages}{178--189} (\bibinfo{year}{1978}).

\bibitem{venkat2022unified}
\bibinfo{author}{Venkatasubramanian, V.}, \bibinfo{author}{Sivaram, A.} \& \bibinfo{author}{Das, L.}
\newblock \bibinfo{title}{A unified theory of emergent equilibrium phenomena in active and passive matter}.
\newblock \emph{\bibinfo{journal}{Computers \& Chemical Engineering}} \textbf{\bibinfo{volume}{164}}, \bibinfo{pages}{107887} (\bibinfo{year}{2022}).

\end{thebibliography}

\section{Supplemental Information}

In this section, we provide six tables that display the layer-by-layer summary of the lognormal parameters for all six networks. 

\includepdf[pages=-, scale=1, offset=0cm -1cm]{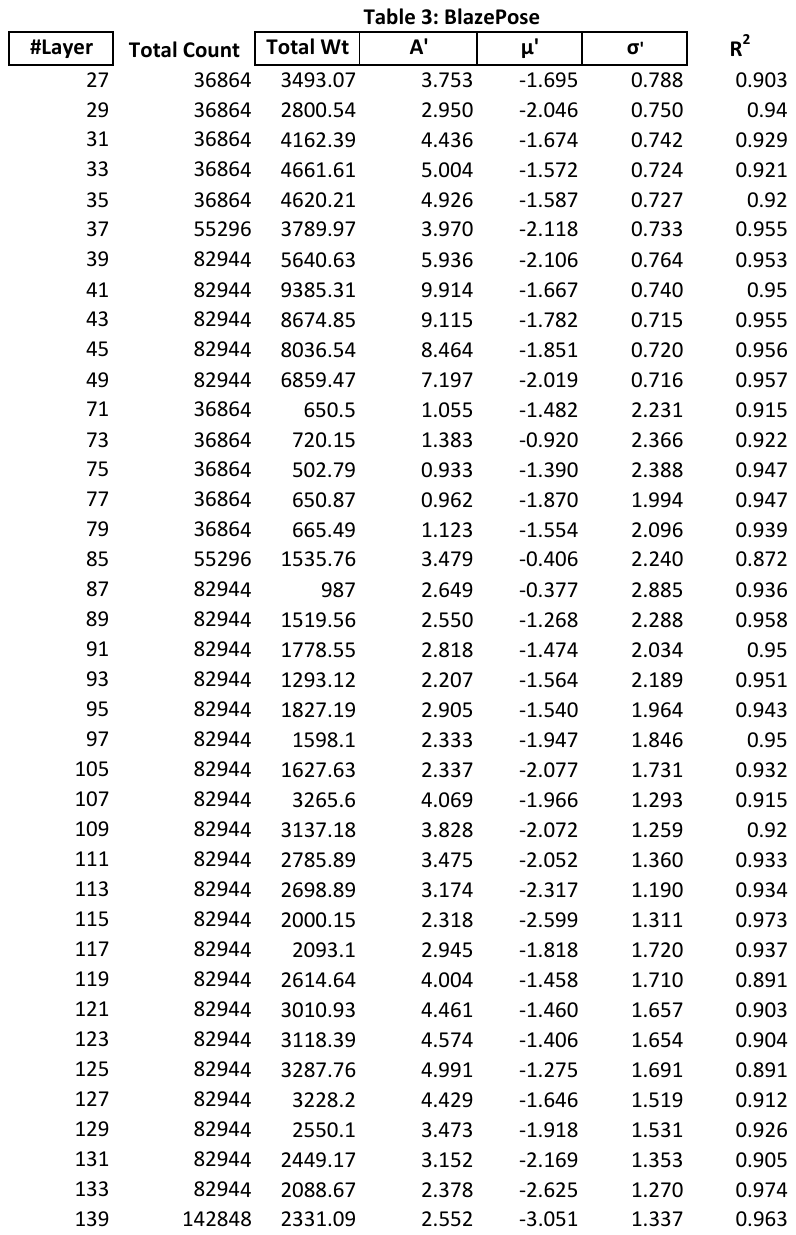}
%
\includepdf[pages=-, scale=1, offset=0cm -1cm]{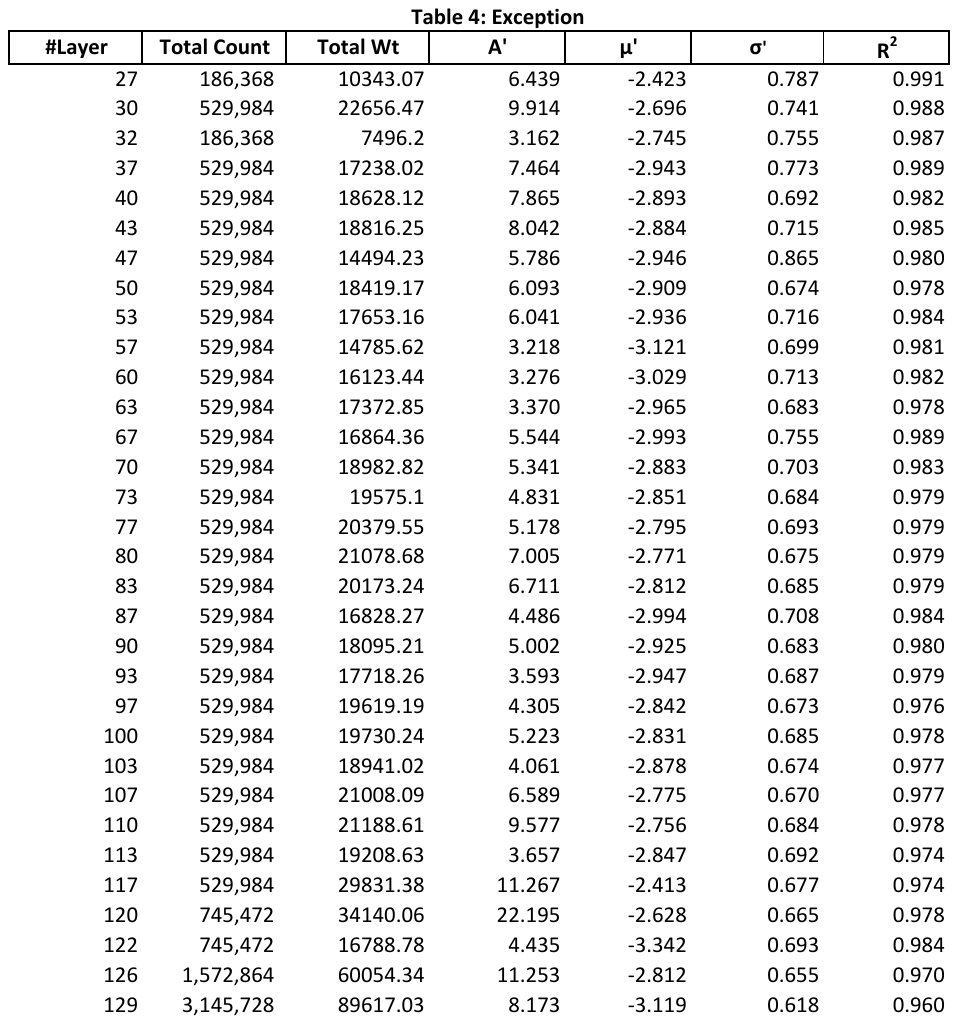}
%
\includepdf[pages=-, scale=1]{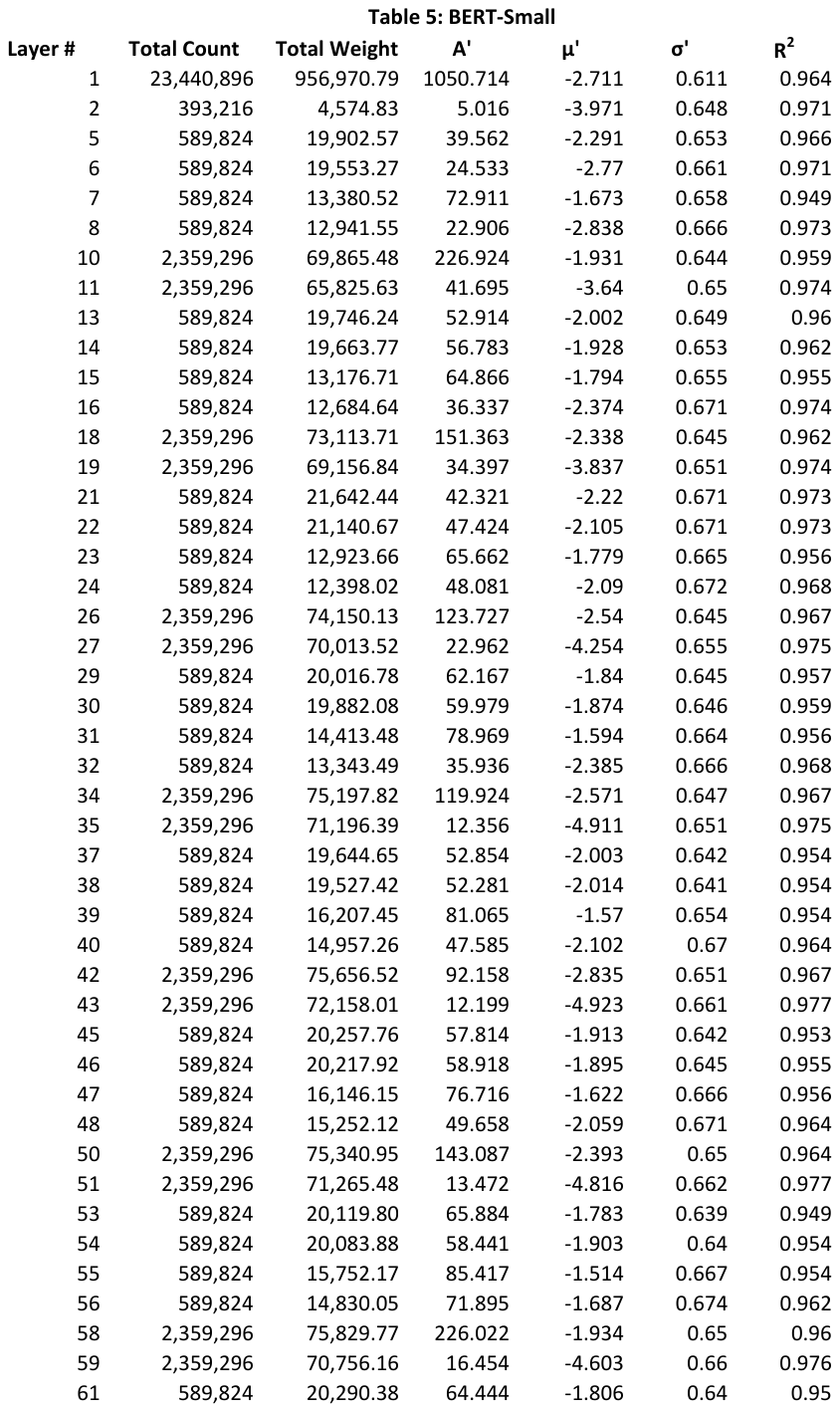}
%
\includepdf[pages=-, scale=1]{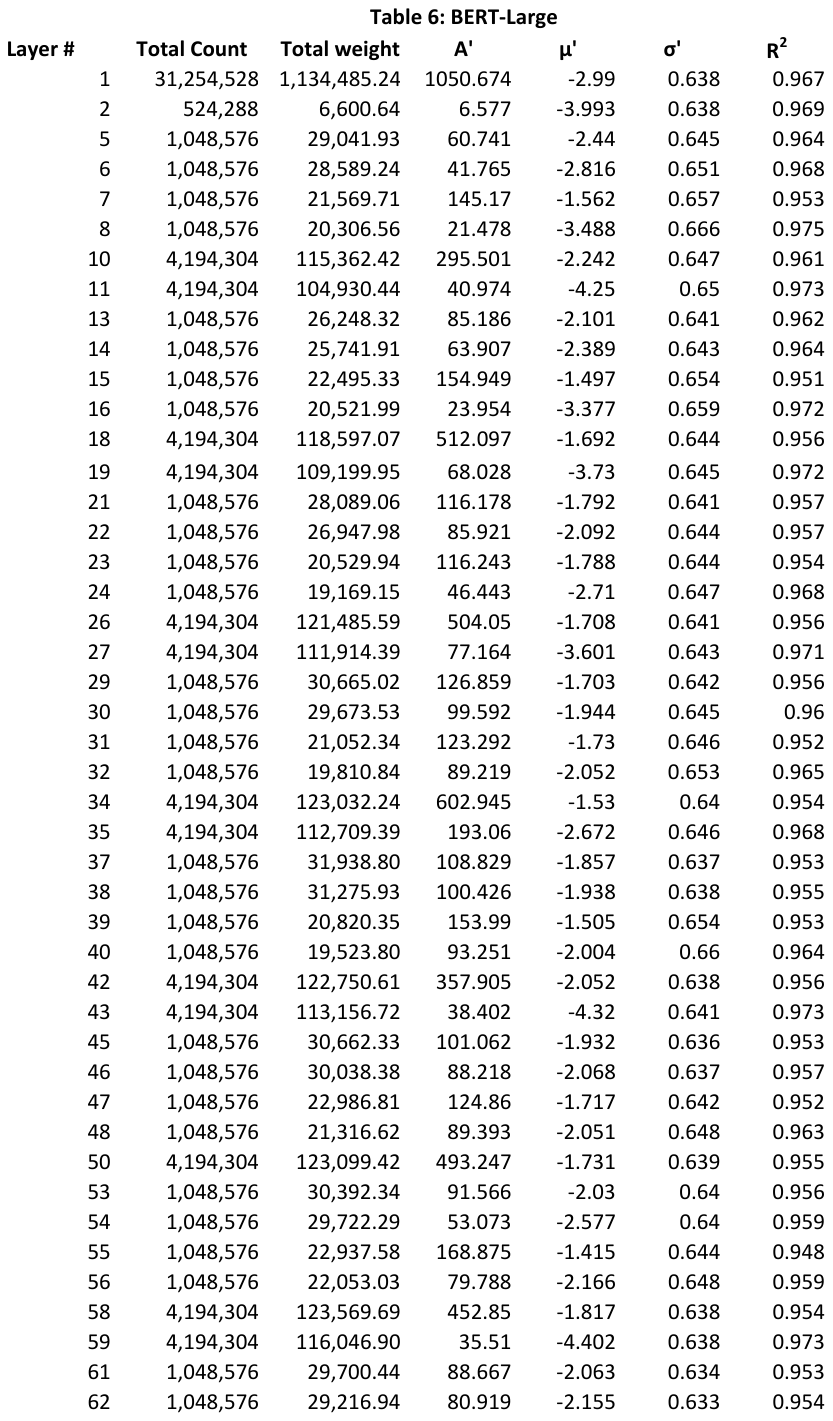}
%
\includepdf[pages=-, scale=1, offset=0cm -1cm]{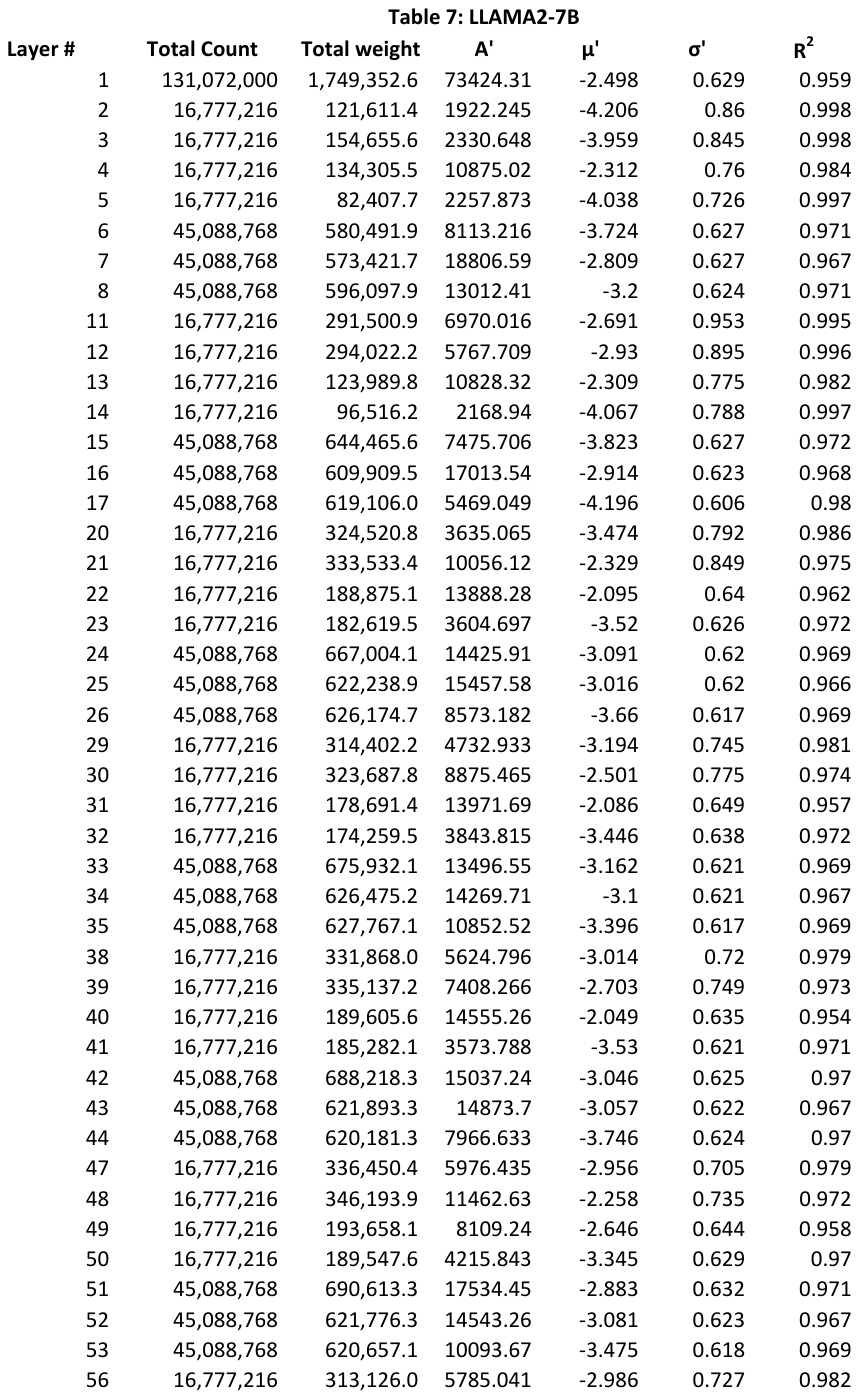}
%
\includepdf[pages=-, scale=1, offset=0cm -1cm]{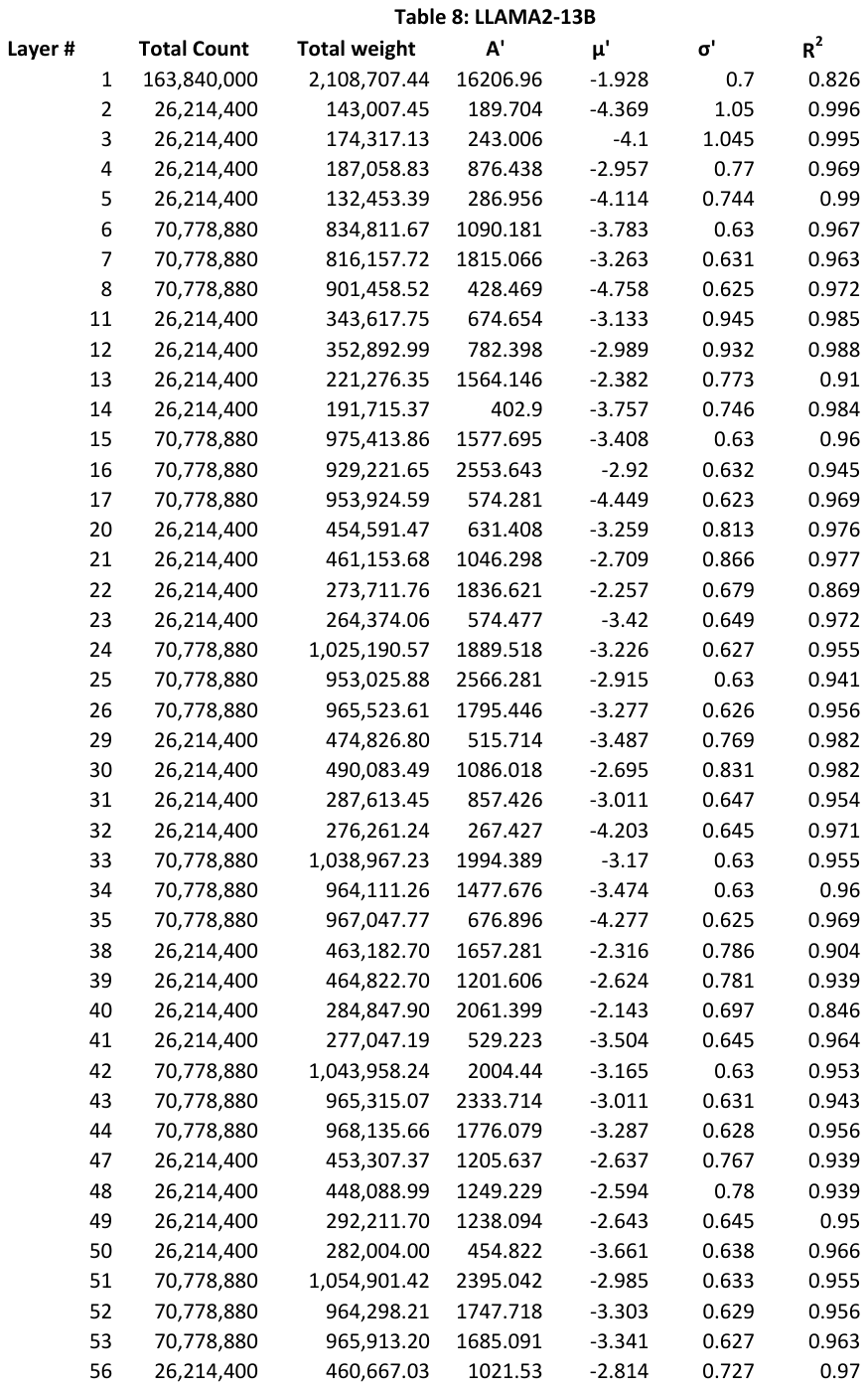}

\end{document}